\documentclass[preprint,showpacs,preprintnumbers,amsmath,amssymb,nofootinbib]{revtex4}
\usepackage{graphicx}
\usepackage{dcolumn}
\usepackage{bm}

\begin{document}
\title{\bf Improved spacecraft radio science using an on-board atomic clock:  \\
application to gravitational wave searches}

\author{Massimo Tinto}
\email{Massimo.Tinto@jpl.nasa.gov} \affiliation{Jet Propulsion
  Laboratory, California Institute of Technology, Pasadena, CA 91109}

\author{George J. Dick}
\email{George.J.Dick@jpl.nasa.gov}
\affiliation{Jet Propulsion Laboratory, California Institute of Technology,
 Pasadena, CA 91109}

\author{John D. Prestage} \email{John.D.Prestage@jpl.nasa.gov}
\affiliation{Jet Propulsion Laboratory, California Institute of
  Technology, Pasadena, CA 91109}
  
\author{J.W. Armstrong}
\email{John.W.Armstrong@jpl.nasa.gov}
\affiliation{Jet Propulsion Laboratory, California Institute of Technology,
 Pasadena, CA 91109}

\date{\today} 
\begin{abstract}
  
  Recent advances in space-qualified atomic clocks (low-mass, low
  power-consumption, frequency stability comparable to that of
  ground-based clocks) can enable interplanetary spacecraft radio
  science experiments at unprecedented Doppler sensitivities.  The
  addition of an on-board digital receiver would allow the up- and
  down-link Doppler frequencies to be measured separately.  Such
  separate, high-quality measurements allow optimal data combinations
  that suppress the currently-leading noise sources: phase
  scintillation noise from the Earth's atmosphere and Doppler noise
  caused by mechanical vibrations of the ground antenna.  Here we
  provide a general expression for the optimal combination of ground
  and on-board Doppler data and compute the sensitivity such a system
  would have to low-frequency gravitational waves (GWs).  Assuming a
  plasma scintillation noise calibration comparable to that already
  demonstrated with the multi-link CASSINI radio system, the
  space-clock/digital-receiver instrumentation enhancements would give
  GW strain sensitivity of $2.0 \times 10^{-17}$ for randomly
  polarized, monochromatic GW signals over a two-decade
  ($\sim0.0001-0.01$ Hz) region of the low-frequency band.  This is
  about an order of magnitude better than currently achieved with
  traditional two-way coherent Doppler experiments.  The utility of
  optimally combining simultaneous up- and down-link observations is
  not limited to GW searches.  The Doppler tracking technique
  discussed here could be performed at minimal incremental cost to
  also improve other radio science experiments (i.e. tests of
  relativistic gravity, planetary and satellite gravity field
  measurements, atmospheric and ring occultations) on future
  interplanetary missions.

\end{abstract}

\maketitle

\section{Introduction}
\label{SecI}

Measurements of the relative velocity between the Earth and an
interplanetary spacecraft, by means of coherent microwave tracking,
have allowed studies of solar system bodies \cite{kliore_etal2004},
tests of relativistic gravity \cite{Vessot1993,bertotti_etal2003},
searches for low-frequency gravitational radiation
\cite{EW1975,Tinto1996,Armstrong2006}, and other science objectives
\cite{kliore_etal2004}. In the frequency band ($10^{-6} - 10^{-2}$)
Hz, typical deep space tracks are limited by phase scintillation
caused by random refractivity variations in the solar wind, and the
ionosphere \cite{asmar_etal2005}.  The most sensitive deep-space
Doppler observations to date, however, calibrate and largely remove
these noises
\cite{bertotti_etal1993,Tinto2002,bertotti_etal2003,armstrong_etal2003}
and are then limited by antenna mechanical noise (unmodeled motion of
the phase center of the ground antenna) and residual post-calibration
tropospheric scintillation (i.e. Doppler fluctuations caused by
refractive index fluctuations in the Earth's atmosphere)
\cite{Tinto1996,asmar_etal2005}. The most sensitive observations hit
the limit identified by these noise sources with an Allan standard
deviation of about $3 \times 10^{-15}$ for integration times of a few
thousand seconds.  Improved sensitivity would benefit the science
disciplines listed above, but antenna mechanical noise, in particular,
has seemed irreducible at reasonable cost since it would require a
large, moving, steel structure much more rigid than that of the
current ground tracking stations.

Several ideas have been proposed to reduce the antenna mechanical
noise or the tropospheric noise, or both
\cite{Armstrong_Estabrook_etal}.  Those ideas do not involve
modifications to the spacecraft, but rather use simultaneous tracking
on the ground with appropriate linear combinations of those data to
synthesize an observable with the ground/tropospheric noises of the
{\it better} of the two receiving stations.  If, however, additional
microwave instrumentation on the spacecraft is considered, such as a
space-borne, high-stability frequency standard
\cite{Prestage_Weaver2007,Dick1990} and a space-qualified digital
receiver \cite{Tyler_etal_2008, RASSI_2008}, then an alternative
method for noise suppression is possible. This involves properly
combining the two one-way (spacecraft to Earth and simultaneous Earth
to spacecraft) Doppler data taken onboard and on the ground in such a
way to maximize the signal-to-noise ratio of the observed physical
observable.\footnote{It should be emphasized that the tropospheric,
  antenna mechanical and ionospheric noise suppression can also be
  accomplished by combining the two-way coherent Doppler data with the
  one-way Doppler measurement performed at the ground
  \cite{Tinto1996,Piran_etal1986}. We have however analyzed the
  configuration with the two one-way measurements for reasons of
  symmetry and simplicity.}  Multiple Doppler observations with
different noises, or different transfer functions to the same noises,
are clearly useful in identifying noise sources and minimizing (and in
some cases canceling) their effects on the final observable.  In
particular, the use of multiple radio links (some of which driven by a
high-quality space-borne frequency standard) was pioneered by R.
Vessot in the Gravity Probe A sub-orbital experiment
\cite{Vessot1970}. Because of mass and power constraints, however, no
very high quality frequency standards have yet been flown on deep
space probes.

Recent advances in clock technology indicate that a new era of
space-qualified, highly stable frequency standards has started
\cite{Prestage_Weaver2007}, which will result into significantly
improved Doppler radio science experiments. A summary of this paper is
given below.

In Section \ref{SecII} we present a brief overview of the theory
underlying Doppler tracking experiments relying only on the two
one-way Doppler data measured onboard and on the ground.  Although
this experimental configuration has been discussed in previous
publications \cite{Vessot1970,Piran_etal1986}, it has been shown only
relatively recently \cite{Tinto1996,Tinto2002} how to fully take
advantage of it for improving the precision of Doppler tracking radio
science experiments. A brief description of an onboard atomic clock
and a digital receiver is given in the Appendix, where an account of
all the one-sided power spectral densities of the noises affecting the
Doppler data is also presented.

The main advantage of spacecraft Doppler tracking experiments relying
on the two one-way Doppler data, over those based on two-way coherent
measurements, is in their ability of exactly canceling the frequency
fluctuations due to the Earth atmosphere and ionosphere, and the
mechanical vibrations of the ground antenna, presently the main noise
sources of Doppler tracking experiments \cite{Tinto1996}. This is
possible because there exists a unique linear combination of the
properly time-shifted one-way measurements that does exactly that
\cite{Tinto1996}. Depending on the specific radio science experiment
performed with this technique, it is actually possible to combine
optimally the two one-way Doppler measurements to maximize the
signal-to-noise ratio (SNR) of the experiment. After deriving the
expression of the optimal SNR in Section \ref{SecIII}, we apply it to
searches for gravitational waves. Under the assumption of calibrating
the frequency fluctuations induced by the interplanetary plasma, we
find that a Doppler broad-band sensitivity of $2.0 \times 10^{-17}$ to
randomly polarized monochromatic signals uniformly distributed over
the sky can be achieved. This is about one order of magnitude better
than that obtainable with two-way coherent Doppler tracking
experiments. Narrow-band searches at frequencies where the transfer
function of the onboard clock reaches sharp nulls (i.e. the
``xylophone'' frequencies) \cite{Tinto1996} further enhance the strain
sensitivity of these experiments to about $7.0 \times 10^{-18}$.

In Section \ref{SecIV} we finally present our comments and
conclusions, and emphasize that the Doppler tracking technique
discussed in this article can be performed at minimal additional cost
by forthcoming interplanetary missions.

\section{The One-Way Doppler Tracking Observables}
\label{SecII}

In Doppler tracking experiments aimed at detecting low-frequency
(milliHertz) gravitational radiation, a distant interplanetary
spacecraft is monitored from Earth through a radio link, with the
Earth and the spacecraft acting as free-falling test particles
\footnote{Spacecraft Doppler GW searches piggy-back on interplanetary
  probes used primarily for other (e.g., solar system) science goals.
  Doppler tracking is the current generation GW detector in the
  low-frequency band.  A much more sensitive, dedicated GW mission,
  {\it LISA}, is currently in the design and technology development
  stage and could launch sometimes in the next decade.}.  A radio
signal of nominal frequency $\nu_0$ is transmitted to the spacecraft,
and coherently transponded back to Earth where the received signal is
compared to a signal referenced to a highly stable clock (typically a
hydrogen-maser).  Relative frequency changes $\Delta \nu / \nu_0$, as
functions of time, are measured.  When a gravitational wave crossing
the solar system propagates through the radio link, it causes small
perturbations in $\Delta \nu / \nu_0$, which are replicated three
times in the Doppler data with maximum spacing given by the two-way
light propagation time between the Earth and the spacecraft
\cite{EW1975}.

\begin{figure}
  \begin{center}
    \includegraphics[width=6.5in, angle = 0.0]{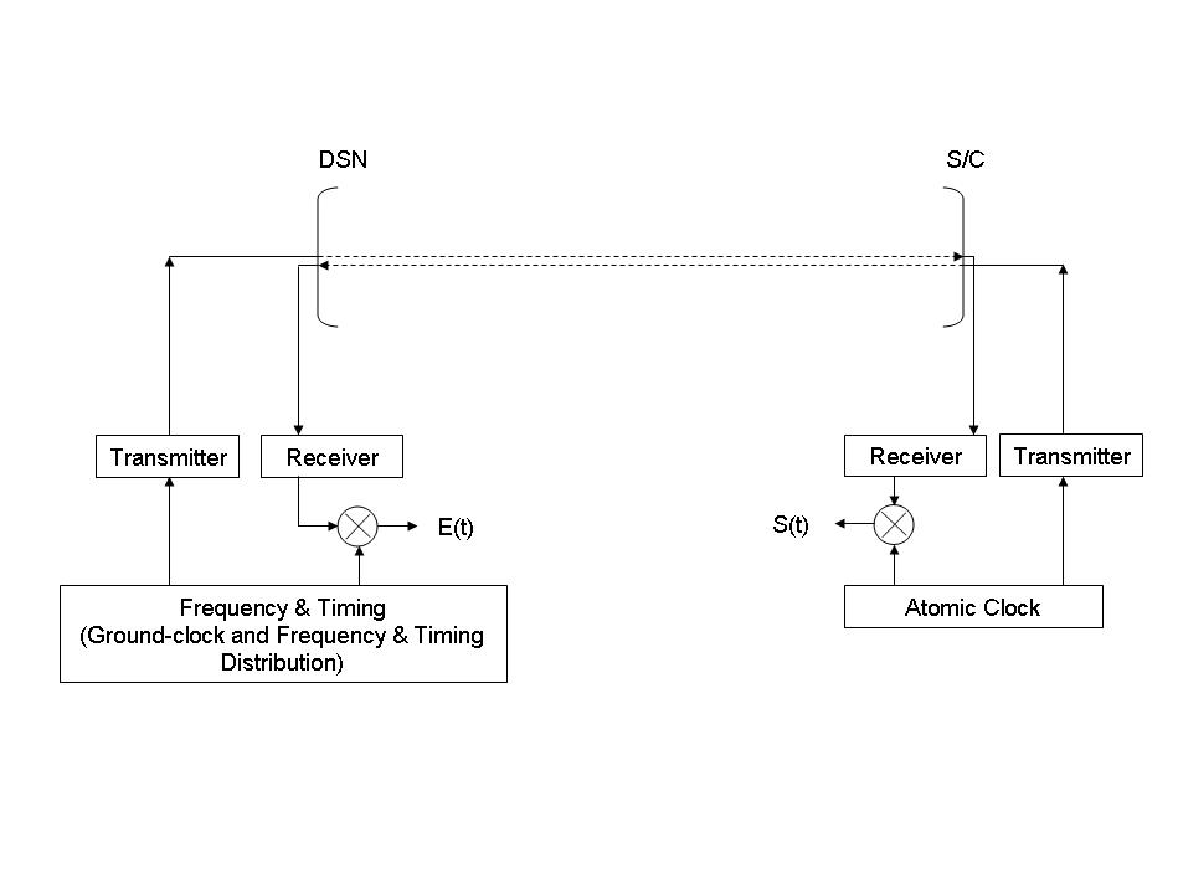}
    \end{center}
    \caption{Block diagram of the radio hardware at the ground
      antenna of the NASA Deep Space Network (DSN) and on board the
      spacecraft (S/C), which allows the acquisition and recording of
      the two Doppler data $E(t)$, $S(t)$. A description of each
      individual block in this diagram is provided in the Appendix.}
\label{Fig1}
\end{figure}

An alternative way of performing Doppler tracking searches for
gravitational radiation was suggested in
\cite{Vessot_Levine1978,Piran_etal1986,Tinto1996}. By adding a
highly-stable clock and a digital receiver to the spacecraft radio
instrumentation (Figure \ref{Fig1}), two one-way Doppler time series
can be recorded simultaneously at the ground station and on board the
spacecraft.

\begin{figure}
  \begin{center}
    \includegraphics[width=6.5in, angle = 0.0]{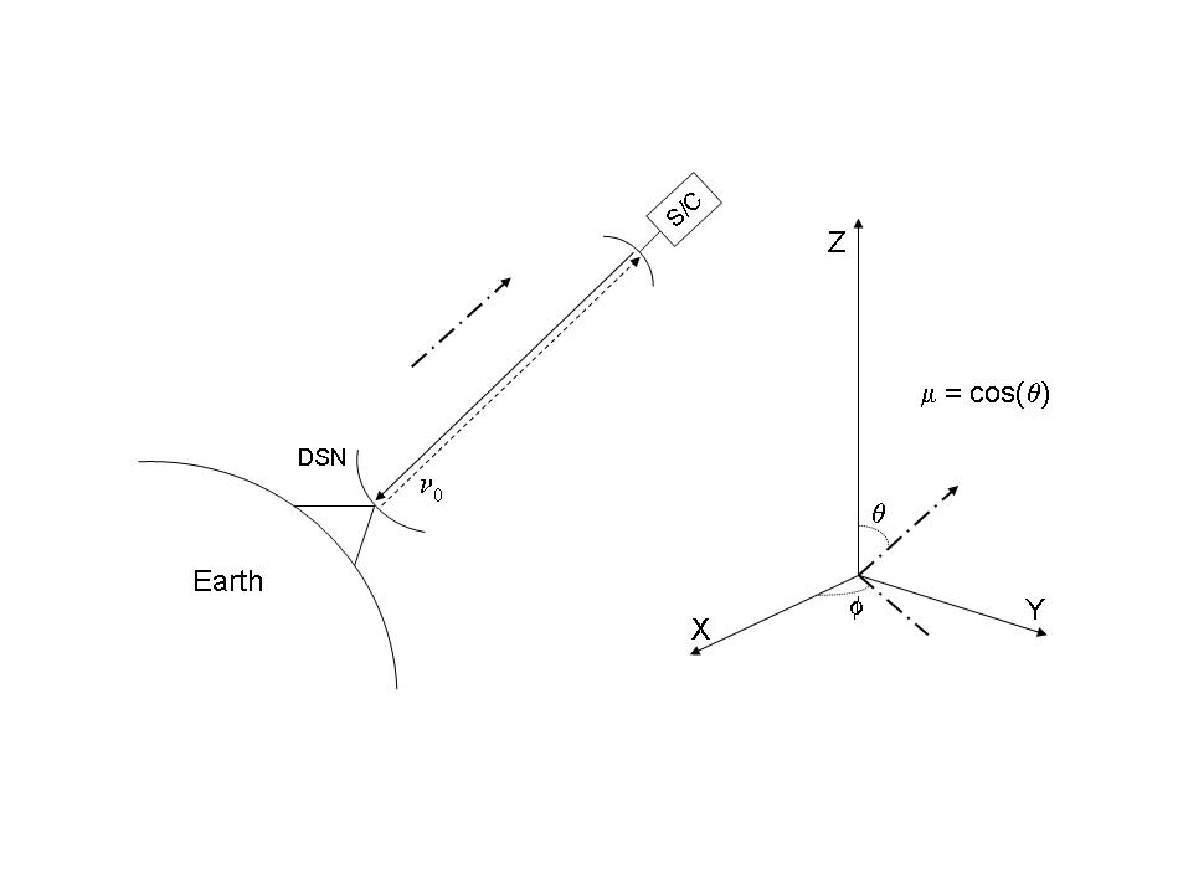}
    \end{center}
    \caption{A radio signal of nominal frequency $\nu_0$ is 
      transmitted to a spacecraft and simultaneously another radio
      signal from the spacecraft and referenced to the onboard clock
      is transmitted to the ground.  The gravitational wave train
      propagates along the $Z$ direction, and the cosine of the angle
      between its direction of propagation and the radio beam is
      denoted by $\mu$. See text for a complete description.}
\label{Fig2}
\end{figure}

If we introduce a set of Cartesian orthogonal coordinates ($X, Y, Z$)
in which the wave is propagating along the $Z$-axis and ($X, Y$) are
two orthogonal axes in the plane of the wave (see Figure \ref{Fig2}),
then the two one-way relative frequency fluctuations at time $t$ can
be written in the following form after first-order Doppler and other
systematic Doppler effects are modeled out from the data
\footnote{The one-way Doppler data measured onboard can be digitally
  recorded, time tagged, and telemetered back to Earth in real time or
  at a later time during the mission.}\cite{Tinto1996}

\begin{eqnarray}
\left(\frac{\Delta \nu (t)}{\nu_0}\right)_{E} \equiv E (t) & = &  
\frac{1 - \mu}{2} \ \left[ h(t - (1 + \mu)L) \ - \ 
h(t) \right] \ + \ C_S (t - L) \ - \ C_{E} (t)  \nonumber \\
& + & T (t) \ + \ B (t - L) \ + \  A_S (t - L) 
\ + \  EL_{E} (t) \ + \ P_{E} (t) \ , 
\label{eq:1} 
\end{eqnarray}
\begin{eqnarray} 
\left(\frac{\Delta \nu (t)}{\nu_0}\right)_{S} \equiv
S (t) & = & \frac{1 + \mu}{2} \ \left[ h(t - L) \ - \ h(t - \mu L)
\right] \ + \ C_{E} (t - L) \ - \ C_S (t)  \nonumber \\
& + & T (t - L) \ + \ B (t) \ + \ 
A_{E} (t - L) \ + \  EL_{S} (t) \ + \ P_{S} (t) \ , 
\label{eq:2} 
\end{eqnarray}
where $h(t)$ is equal to 
\begin{equation} 
h (t) = h_+(t) \cos (2 \phi) + h_{\times}(t) \sin (2 \phi) \ .
\end{equation}
Here $h_+(t)$, $h_\times (t)$ are the wave's two amplitudes with
respect to the ($X, Y$) axis, ($\theta, \phi$) are the polar angles
describing the location of the spacecraft with respect to the ($X, Y,
Z$) coordinates, $\mu$ is equal to $\cos \theta$, and $L$ is the
distance to the spacecraft (units in which the speed of light $c = 1$)

In Eqs. (\ref{eq:1}, \ref{eq:2}) we have assumed the Earth and the on
board clocks to be perfectly synchronized. Although this condition is
impossible to achieve in practice, it has been previously shown by one
of us \cite{Tinto2002} that the accuracy required for successfully
implementing a noise cancellation scheme similar to the one discussed
in this paper requires a clock synchronization accuracy of about $0.5
\ s$, which is easy to achieve.

In Eqs. (\ref{eq:1},\ref{eq:2}) we have denoted by $C_E (t)$, $C_S
(t)$ the random processes associated with the frequency fluctuations
of the clock on Earth and onboard respectively, $B(t)$ the joint
effect of the noise from buffeting of the probe by non gravitational
forces and from the antenna of the spacecraft, $T(t)$ the joint
frequency fluctuations due to the troposphere, ionosphere and ground
antenna, $A_E(t)$ the noise of the radio transmitter on the ground,
$A_S(t)$ the noise of the radio transmitter on board, $EL_{E} (t)$,
$EL_{S} (t)$, the noise from the electronics on the ground and onboard
respectively, and $P_{E}(t)$, $P_{S}(t)$ the fluctuations on the two
links due to the interplanetary plasma. Since the frequency
fluctuations induced by the plasma are, to first order, inversely
proportional to the square of the radio frequency, by using high
frequency radio signals or by monitoring two different radio
frequencies transmitted to and from the spacecraft, this noise source
can be suppressed to very low levels or entirely removed from the data
respectively \cite{bertotti_etal1993}. In what follows we will assume
dual frequency to be used, and disregard the noise effects of the
plasma fluctuations in our analysis.

From Eqs. (\ref{eq:1},\ref{eq:2}) we deduce that gravitational wave
pulses of duration longer than the one-light-time $L$ give a Doppler
response that, to first order, tends to zero.  The tracking system
essentially acts as a pass-band device, in which the low-frequency
limit $f_l$ is roughly equal to $(L)^{-1}$ Hz, and the high-frequency
limit $f_H$ is set by the thermal noise in the receiver.  Since the
clocks and some electronic components are most stable at integration
times around $1000$ seconds, Doppler tracking experiments are
performed when the distance to the spacecraft is of the order of a few
astronomical units.  This sets the value of $f_l$ for a typical
experiment to about $10^{-4}$ Hz, while the thermal noise gives an
$f_H$ of about $10^{-2}$ Hz.

It is important to note the characteristic time signatures of the
clock noises, $C_E (t)$ and $C_S (t)$, of the probe antenna and
spacecraft buffeting noise $B(t)$, of the troposphere, ionosphere, and
ground antenna noise $T(t)$, and the transmitters $A_E(t)$, $A_S(t)$.
The time signature of the two clock noises, for instance, can be
understood by observing that the frequency of the signal received at
the ground station at time $t$ contains fluctuations from the onboard
clock that were transmitted $L$ seconds earlier and the noise from the
ground clock enters with a negative sign at time $t$ due to the
heterodyne nature of the Doppler measurement. The time signature of
the noises $T$, $B(t)$, $A_E(t)$, and $A_S(t)$ in Eq.
(\ref{eq:1},\ref{eq:2}) can be understood through similar
considerations.

Since the major noise source affecting the two one-way measurements is
represented by the fluctuations induced by the Earth troposphere and
the mechanical vibrations of the ground station, it has been
emphasized \cite{Tinto1996} that there exists a combination of the two
Doppler data that cancels these noises. It is easy to see from
inspection of Eqs. (\ref{eq:1},\ref{eq:2}) that such a combination is
equal to
\begin{equation}
x (t) \equiv S(t) - E(t - L) \ .
\label{x}
\end{equation}
After substituting into Eq. (\ref{x}) the expressions for $E(t)$,
$S(t)$ given in Eqs. (\ref{eq:1},\ref{eq:2}) we get
\begin{eqnarray}
x (t) & = & h (t - L) - \frac{1 + \mu}{2} \ h(t - \mu L) - \frac{1 -
  \mu}{2} \ h(t - 2L - \mu L) 
\nonumber
\\
&& + \ 2 C_E (t - L) - [C_S(t) + C_S(t - 2L)] + [B(t) - B(t - 2L)]
\nonumber
\\
&& + \ A_E (t - L) - A_S (t - 2L) + EL_S (t) - EL_E (t - L) \ .
\label{eq:3}
\end{eqnarray}

From Eq. (\ref{eq:3}) we may notice that the spacecraft buffeting
noise, $B$, does not cancel exactly and it gets suppressed by its
transfer function to the $x$ combination at frequencies smaller than
the inverse of the round-trip-light-time.

\section{Gravitational Wave Sensitivities}
\label{SecIII}

This section describes the derivation of the sensitivity, which is
defined on average over the sky, to be equal to the strength of a
sinusoidal gravitational wave required to achieve a signal-to-noise
ratio of $1$ in a forty-day integration time. Note that the
sensitivity is therefore a function of Fourier frequency, $f$. We have
chosen the integration time to be equal to forty days since this was
the tracking time of the CASSINI gravitational wave experiments
\cite{Armstrong2006}.  Sensitivity is essentially the noise-to-signal
ratio and it will be computed for both the new data combination $x$ as
well as for the traditional two-way coherent tracking measurement,
$y$, for comparison reasons. For convenience we provide below the
expression of the two-way Doppler response, $y(t)$, which will be used
later on for estimating its sensitivity

\begin{eqnarray}
y (t) & = & - \frac{(1 - \mu)}{2} \ h(t) \ - \ \mu \ h(t - (1 + \mu)L) + 
\frac{(1 + \mu)}{2} \ h(t - 2L) 
\nonumber 
\\ 
& + & C_E (t - 2L) \ - \ C_E (t) \ + \ 2B (t - L) \ + 
\ T (t - 2L) \ + \ T (t) 
\nonumber
\\ 
& + & A_E (t - 2L) \ + \ A_S (t - L) \ 
\ + \ TR (t - L) \ + \ EL_{E_2} (t) \ ,
\label{eq:4}
\end{eqnarray} 
where $TR$ is the random process associated with the relative
frequency fluctuations due to the onboard microwave transponder. For
more details we refer the reader to \cite{Armstrong1987}.

\subsection{Signal Averaged Power}

The averaged signal power in the combination $x$, estimated at an
arbitrary Fourier frequency $f$, is computed by (i) taking the Fourier
transform of the signal entering into the combination $x$ and its
modulus-squared, and (ii) by integrating the resulting expression over
an ensemble of sinusoidal signals uniformly distributed over the
celestial sphere. Such a calculation is long but straightforward, and
the resulting expression, $S_{x_h}$, is equal to
\begin{equation}
S_{x_{h}} = S_h \ \left[ \frac{4}{3} + \frac{2}{3} \ \cos^2(2 \pi f L) -
\frac{\sin^2(2 \pi f L)}{2 (\pi f L)^2} \right] \ ,
\label{Sx}
\end{equation}
where $S_h$ is the gravitational wave signal one-sided power spectral
density.

The calculation of the averaged signal power of the observable $y$ can
similarly be carried through, resulting into the following expression
\begin{equation}
S_{y_h} = S_h \ 
\frac{8 \pi^3 f^3 L^3 + 2 \sin(4 \pi f L) 
- \frac{2}{3} \pi f L \ (3 + 4 \pi^2 f^2 L^2) \cos(4 \pi f L) -6 \pi f L}{8 \pi^3 f^3 L^3}
\label{Sy}
\end{equation}

\begin{figure}
  \begin{center}
    \includegraphics[width=6.5in, angle = 0.0]{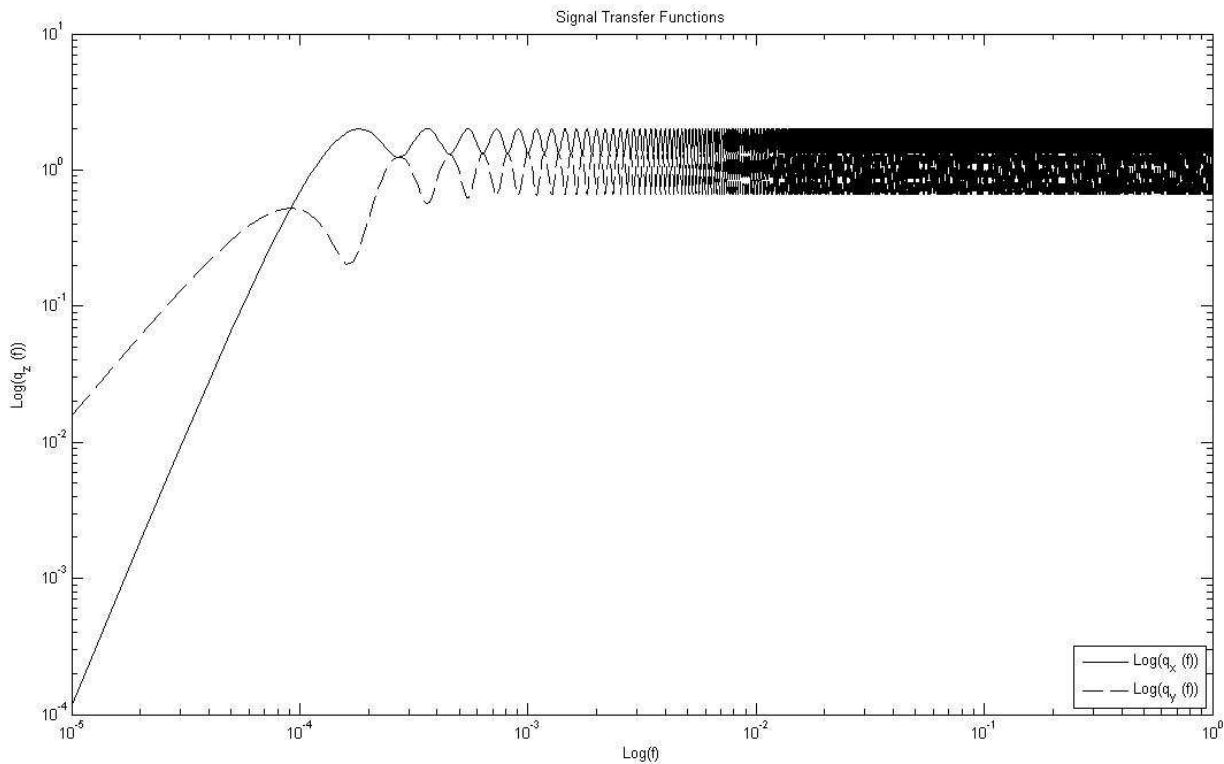}
    \end{center}
    \caption{Averaged power transfer functions of the Doppler
      responses $x$ and $y$ to an ensemble of sinusoidal signals
      randomly polarized and uniformly distributed over the celestial
      sphere. The x-transfer function shows constructing interference
      at frequencies that are integer multiples of the inverse of the
      round-trip-light-time, $2L$, taken here to be equal to $5500$
      seconds. The coupling of the $x$ data combination to such a
      stochastic ensemble of gravitational radiation is slightly
      stronger than that of the two-way $y$ response at frequencies
      larger than $5.0 \times 10^{-4}$ Hz. At lower frequencies the
      transfer function of the $x$ combination decays more rapidly
      than that of $y$ as a consequence of being the difference of the
      two one-way measurements.}
\label{Fig3}
\end{figure}

Figure \ref{Fig3} shows the two ``transfer functions'' of the signal
one-sided power spectral densities, i.e. $ q_{x} \equiv S_{x_{h}}/S_h$
and $q_y \equiv S_{y_h}/S_h$. The transfer function $q_{x}$ is
slightly larger than $q_y$ in the region $[5 \times 10^{-4} - 1]$ Hz,
indicating that a constructing interference of the signal with itself
is taking place in this part of frequency band. On the other hand, in
the low-part of the frequency band the combination $x$ shows a
coupling to gravitational radiation that is weaker than that of $y$.
This is to be expected, as $x$ is the difference of the two one-way
measurements, which in the ``long-wavelength'' limit become equal to
each other.

\subsection{Noise Spectra}

To compute the sensitivity of the combination $x$, and compare it
against that of the two-way Doppler measurement, $y$, we need the
one-sided power spectral densities of the main noise sources and their
transfer functions to the observables $x$ and $y$. If we assume all
the noise sources to be uncorrelated, from equations (\ref{eq:3},
\ref{eq:4}) we can derive the following expressions for the two noise
spectra $S_{x_n}$ and $S_{y_n}$
\begin{eqnarray}
S_{x_n} & = & 4 S_{C_E} + 4 S_{C_S} \cos^2(2 \pi f L) + 4 S_{B}
\sin^2(2 \pi f L) + S_{A_{E}} + S_{A_S} + S_{EL_{E}} + S_{EL_S}
\ ,
\label{Snoise_x}
\\
S_{y_n} & = & 4 S_{C_E} \sin^2(2 \pi f L) + 4 S_{T} \cos^2(2 \pi f L)
+ 4 S_{B} + S_{A_{E}} + S_{A_S} + S_{EL_{E}} + S_{EL_S} + S_{TR} \ ,
\label{Snoise_y}
\end{eqnarray}
where the meaning of the various terms appearing into Eqs.
(\ref{Snoise_x},\ref{Snoise_y}) is self-explanatory. We provide in the
Appendix the expressions for the various noise spectra corresponding
to a gravitational wave search performed with a spacecraft out to a
distance of $5.5$ AU from Earth (as was during the first gravitational
wave experiment with the CASSINI spacecraft), and equipped with an
onboard microwave instrumentation similar to the one flown on CASSINI.

The gravitational wave sensitivity is the wave amplitude required to
achieve a signal-to-noise ratio of $1$, and it can be computed as a
function of Fourier frequency using the following expression
\cite{Armstrong2006}
\begin{equation}
\Sigma_z (f) \equiv \sqrt{ \frac{S_{z_n}(f) \ B}{q_z (f)}} \ ,
\end{equation}
where $z$ means either $x$ or $y$, and $B$ is the frequency bandwidth
corresponding to a $40$ days integration time, the duration of the
gravitational wave experiments performed with the CASSINI spacecraft
\cite{Armstrong2006}.

\begin{figure}
  \begin{center}
    \includegraphics[width=6.0in, angle = 0.0]{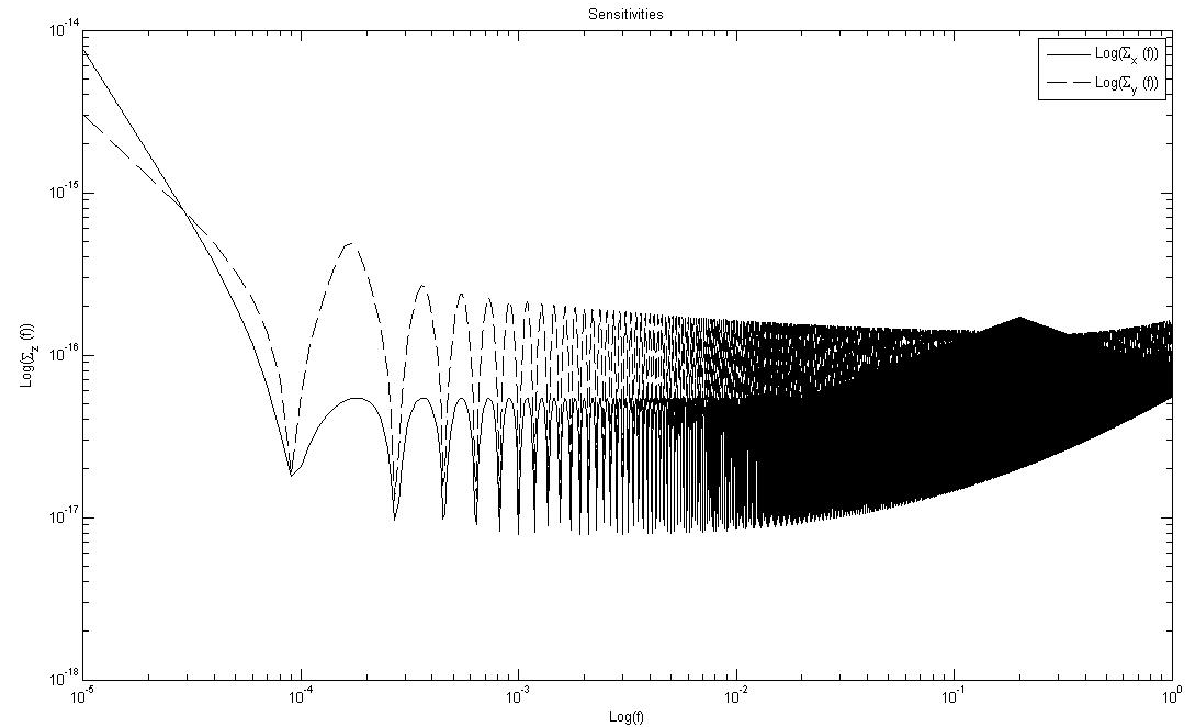}
    \end{center}
    \caption{Sensitivities of the $x$ (solid-line) and $y$ (dash-line) 
      Doppler responses to a randomly polarized and uniformly
      distributed stochastic ensemble of sinusoidal gravitational wave
      signals. The sensitivity is expressed as a function of the
      frequency and represents the equivalent sinusoidal strain
      required to produce a signal-to-noise ration of $1$. The
      spacecraft is assumed to be out to a distance of $5.5$ AU, and
      eighty percent tropospheric noise calibration is applied to the
      two-way Doppler data $y$ (dash-line). The two sensitivity curves
      reflect the noise spectral levels and shapes (given in the
      Appendix), their transfer functions to the observables $x$ and
      $y$ (Eqs.  \ref{Snoise_x}, \ref{Snoise_y}), and the
      gravitational wave transfer functions shown in Fig. \ref{Fig3})}
\label{Fig4}
\end{figure}

The main difference between the two observables $x$ and $y$ is of
course in the absence in the $x$ combination of the joint disturbances
from the troposphere and the mechanical vibration of the ground
antenna (the random process denoted $T$ in Eq.(\ref{eq:4})) and the presence
(in $x$) of the spacecraft clock noise process. As
onboard and ground microwave instrumentation have in recent years
reached unprecedented frequency stabilities, $T$ has indeed become the
major sensitivity limitation of two-way Doppler tracking searches for
gravitational radiation \cite{Armstrong2006}.  Since the effects from
the troposphere can be mitigated by relying on simultaneous
measurements performed by a radiometer located in the proximity of the
tracking station, our sensitivity analysis reflects the assumption of
being able to calibrate out eighty percent of the tropospheric effects
from the $y$ observable (as demonstrated by the CASSINI experiments).
Figure \ref{Fig4} shows the estimated sensitivity of the observable
$x$ (continuous-line) formed out of the two one-way measurements, and
compares it against that of the two-way measurement in which eighty
percent of the tropospheric effects are calibrated out (dash-line).
The $x$ combination displays the best sensitivity in the frequency
band $10^{-4} - 10^{-1}$ Hz, which is of most interest to
gravitational wave search experiments. At higher frequencies, between
about $10^{-1} - 3.0 \times 10^{-1}$ Hz, effects related to the
locking of the atomic clock to its local oscillator introduce a small
degradation in the $x$ sensitivity over that of the $y$ data.  Also,
at frequencies lower than $3.0 \times 10^{-5}$ Hz, the $x$ combination
shows a sensitivity worse than that of $y$ due to the cancellation of
the signal in this low-frequency region.

As the two one-way Doppler measurements can be combined to synthesize
the two-way measurement $y$ \cite{Piran_etal1986,Tinto1996}, one could
argue that at these frequencies one could of course rely on the
synthesized $y$ data to take advantage, if needed, of its better
sensitivity in these two regions of the accessible frequency band.
This observation suggests that it must be possible to identify a
combination of the two one-way Doppler data that maximizes the
sensitivity to gravitational waves in the entire band of interest.

\subsection{Optimal Sensitivity}

In order to derive the combination of the two one-way Doppler data
that achieves optimal sensitivity, let us consider the following
linear combination $\eta (f)$ of the Fourier transforms of $E (t)$ and
$S(t)$
\begin{equation}
\eta(f) \equiv a_1 (f, {\vec \lambda}) \ {\widetilde{E}} (f) \ +
\ a_2 (f, {\vec \lambda}) \ {\widetilde{S}} (f) \ ,
\label{eq:8}
\end{equation}
where the $\{a_i (f, \vec \lambda)\}_{i=1,2}$ are arbitrary complex
functions of the Fourier frequency $f$, and of a vector $\vec \lambda$
containing parameters characterizing the signal and the noises
affecting the two Doppler data.  For a given choice of the two
functions $\{a_i \}_{i=1,2}$, $\eta$ gives a specific Doppler data
combination, and our goal is therefore that of identifying, for a
given signal, the two functions $\{a_i \}_{i=1,2}$ that maximize the
signal-to-noise ratio \cite{Helstrom68}, $SNR_{\eta}^2$, of the
combination $\eta$
\begin{equation}
SNR_{\eta}^2 = 
\int_{f_{l}}^{f_u} 
\frac{|a_1 \ {\widetilde E_{s}} + a_2 \ {\widetilde S_{s}}|^2} 
{{\langle|a_1 \ {\widetilde E_{n}} + a_2 \ {\widetilde S_{n}}|^2
  \rangle}} \ df \ .
\label{eq:9bis}
\end{equation}
In equation (\ref{eq:9bis}) the subscripts $s$ and $n$ refer to the
signal and the noise parts of (${\widetilde {E}}, {\widetilde {S}}$)
respectively, the angle brackets represent noise ensemble averages,
and the interval of integration ($f_l, f_u$) corresponds to the
accessible frequency band.

The $SNR_{\eta}^2$ can be regarded as a functional over the space of
the two complex functions $\{ a_i \}_{i=1,2}$, and their expressions
that maximize it can of course be derived by solving the associated
set of Euler-Lagrange equations. The derivation of the expression of
the optimal SNR, ${SNR_{\eta}^2}_{\rm opt}$, is long but
straightforward, and it is equal to (see \cite{PTLA02} for details)
\begin{equation}
{SNR_{\eta}^2}_{\rm opt.} = \int_{f_{l}}^{f_u} 
{\bf z}^{(s)*}_i \ ({\bf C}^{-1})_{ij} \ {\bf z}^{(s)}_j  \ df \ .
\label{eq:18}
\end{equation}
\noindent
In Eq. (\ref{eq:18}) the convention of sum over repeated indices is
assumed, ${\bf z}^{(s)}$ is the vector of the
signals, (${\widetilde{E_{s}}}, {\widetilde{S_{s}}}$), and
${\bf C}$ is the hermitian, non-singular, correlation matrix of the
vector random process ${\bf z}_n \equiv ({\widetilde{E_{n}}},
{\widetilde{S_{n}}})$
\begin{equation}
({\bf C})_{rt} \equiv \langle {\bf z}^{(n)}_{r} {\bf z}^{(n)*}_{t} \rangle \ .
\label{eq:14}
\end{equation}

\begin{figure}
\begin{center}
\includegraphics[width=6.0in, angle = 0.0]{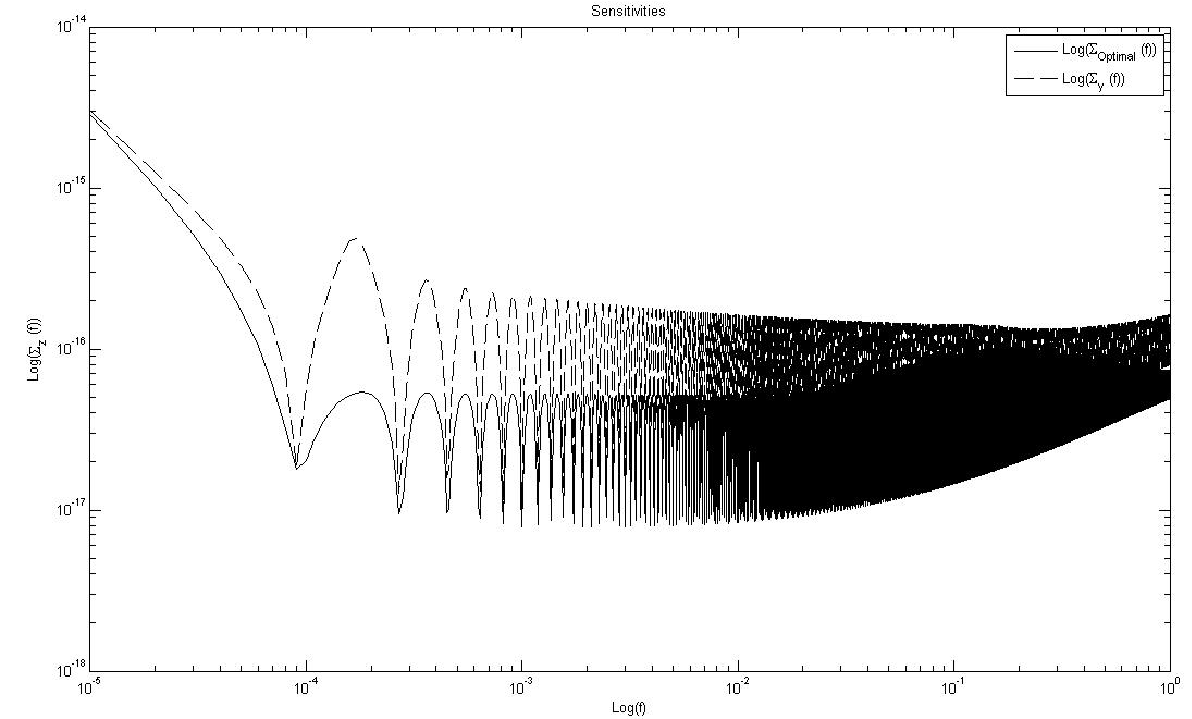}
\end{center}
\caption{Sensitivity curves of the optimal combination (solid-line)
  and the two-way Doppler tracking data (dash-line). The noise
  parameters are equal to those used in figure \ref{Fig4}.  Note how
  the sensitivity of the optimal combination is now consistently below
  that of the two-way measurement.}
\label{Fig5}
\end{figure}

Eq. (\ref{eq:18}) can now be used for estimating the sensitivity to an
ensemble of sinusoidal gravitational wave signals randomly polarized
and uniformly distributed over the celestial sphere. Figure \ref{Fig5}
shows the estimated {\it optimal} sensitivity (solid-line) obtained by
relying on the two one-way measurements, and again that of a two-way
coherent tracking experiment (dash-line), in which all the parameters
characterizing the experiment are as in Figure (\ref{Fig4}).  Note how
the sensitivity of the optimal combination is now consistently below
that of the two-way measurement, and it coincides with that of the $x$
combination in most of the accessible frequency band.

\section{Conclusions}
\label{SecIV}

We have discussed a method for significantly enhancing the sensitivity
of Doppler tracking experiments aimed at the detection of
gravitational waves. The main result of our analysis shows that by
adding an atomic frequency standard and a digital receiver on board
the spacecraft we can achieve a broad-band sensitivity of $2.0 \times
10^{-17}$ in the milliHertz band. This sensitivity figure is obtained
by completely removing the frequency fluctuations due to the
interplanetary plasma. Our method relies on a properly chosen linear
combination of the one-way Doppler data recorded on board with the
data measured on the ground. It allows us to optimally suppress the
frequency fluctuations due to the troposphere, ionosphere, and antenna
mechanical and, for a spacecraft that is tracked for $40$ days out to
$5.5$ AU, to reach a sensitivity that is about one order of magnitude
better than that achievable by a state-of-the art two-way Doppler
tracking search.

The expression of the optimal combination of the two one-way Doppler
data can be used in all the classic tests of relativistic theory
of gravity in which one-way and two-way spacecraft Doppler
measurements are used as primary data sets.  We will analyze the
implications of the sensitivity improvements that this technique will
provide for direct measurements of the gravitational red-shift, the
second-order relativistic Doppler effect predicted by the theory of
special relativity, searches for possible anisotropy in the velocity
of light, measurements of the parameterized post-Newtonian parameters,
and measurements of the deflection and time delay by the Sun in radio
signals. This research is in progress, and will be the subject of a
forthcoming investigation.

\section*{Acknowledgements} 
\label{SecV}

It is a pleasure to that Frank B. Estabrook for his constant
encouragement during the development of this work. This research was
performed at the Jet Propulsion Laboratory, California Institute of
Technology, under contract with the National Aeronautics and Space
Administration.(c) 2008 California Institute of Technology. Government
sponsorship acknowledged.

\section*{Appendix} 
\label{SecVI}

\subsection{Noise sources and their spectra}

This appendix provides a general description of the radio hardware needed for
implementing the technique described in the main body of this paper,
the corresponding one-sided power spectral densities of the frequency
fluctuations introduced by these subsystems into the observables $E
(t)$ and $S (t)$, and discusses the frequency fluctuations due to the
Earth atmosphere. For a more comprehensive analysis on the radio
hardware the reader is referred to
\cite{Tinto1996,Tinto2000,Armstrong2006,Prestage_Weaver2007,Dick1990,RASSI_2008},
while a review on the propagation noises is given in
\cite{Armstrong2006}

The ground {\it master clock} and the {\it frequency \& timing
  distribution} represent the overall contribution of the reference
clock itself and the cabling system that takes the signal generated by
the master clock to the antenna.  This can be located several
kilometers away from the site of the clock, implying that the need of
a highly-stable cabling system is required.  It has been shown at JPL
that optical-fiber cables would not degrade significantly the
frequency stability of the signal generated by the master clock.  The
corresponding one-sided power spectral density of the frequency
fluctuations, introduced by these two noise sources, is equal to
\cite{Tinto2000}
\begin{equation}
S_{C_E} (f) =  6.2 \times [10^{-28} f + 10^{-33} f^{-1} + 10^{-30}] +
1.3 \times 10^{-28} f^2 \ \ {\rm Hz}^{-1} \ .
\end{equation}

The {\it Ground and onboard Transmitters} include all the frequency
multipliers that are needed to generate the desired frequency of the
transmitted radio signal, starting from the frequency provided by the
clocks.  It also accounts for the radio amplifier, and the extra phase
delay changes occurring between the amplifier and the feed cone of the
antennas.  The noise due to the amplifiers is the dominant one, and it
has been characterized in \cite{Tinto1996}.  The one-sided power spectral
densities of the frequency fluctuations are given by
\begin{equation}
S_{A_E} (f) + S_{A_S} (f) = 
2.3 \times 10^{-28} + 4.0 \times 10^{-25} f \ \ {\rm Hz}^{-1} \ ,
\end{equation}

The noises introduced by the {\it Receiver} at the ground station can
be modeled as white phase fluctuations.  The contribution to the
overall noise budget from the receiver chain on the ground can be
repartitioned into thermal noise (finiteness of the signal-to-noise
ratio) and fluctuations introduced into the signal as it propagates
through the cables and waveguides running from the feed of the antenna
to the actual receiver.  The effects of the latter noise source is
nowadays minimized with the use of beam waveguide (BWG) antennas.
These new antennas have become operational in the year 2004 at the
NASA Deep Space Network three sites: in North America (Goldstone,
California), Europe (Madrid, Spain), and Australia (Canberra).

Under the assumption of relying on a $34$ meter diameter beam waveguide
antenna for receiving a coherent Ka-Band ($32$ GHz) signal transmitted by
a spacecraft out a distance of $5.5$ AU, a ground system noise
temperature of about $70$ degrees Kelvin, an onboard Ka-Band amplifier
of $10$ W, and a spacecraft High Gain Antenna (HGA) with a diameter of
about $4$ meters, we find the following one-sided power spectral
density of the frequency fluctuations at Ka-Band \cite{Armstrong2006}
\begin{equation}
S_{EL_E} (f) = 6.3 \times 10^{-27} \ f^2 \ {\rm Hz}^{-1} \ .
\end{equation}
The {\it buffeting} of the spacecraft will introduce unwanted
frequency fluctuations in the one-way Doppler observable.  Estimates
of its magnitude have been given in \cite{Riley_etal1990}, and the
one-sided power spectral density of the frequency fluctuations is
given by the following expression
\begin{equation}
S_B (f) = 5.0 \ \times 10^{-42} \ f^{-3} \ + \  1.0 \ \times 10^{-31}
\ {\rm Hz}^{-1} \ .
\end{equation}
The noise introduced into the Doppler observable $y$ by the Earth {\it
  Atmosphere and ionosphere}, and by the scintillation of the {\it
  interplanetary plasma}, have been studied extensively in the
literature \cite{AWE1979,Armstrong2006,Keihm1995}.

The scintillation introduced into the Doppler observables by the {\it
  Atmosphere} are independent of the microwave frequency at which the
spacecraft is tracked.  Since gravitational wave searches
are performed in a band whose upper frequency cutoff is smaller than
$1$ Hz (thermal noise at higher frequencies becomes unacceptably
large), the one-sided power spectral density of the noise due to the
atmosphere can be written as follows \cite{Linfield1998}

\begin{eqnarray}
S_T (f) & = & 2.8 \ \times 10^{-28} \ f^{-2/5} \ \ {\rm Hz}^{-1} \ \ \ \ \ \ \
10^{-5} \le f \le 10^{-2} \ \ {\rm Hz} \nonumber \\ 
& = & 2.2 \ \times 10^{-30} \ f^{-3} \ \ \ \ {\rm Hz}^{-1} \ \ \ \ \ \ \
10^{-2} \le f \le 1  \ \ \ \ \ \ \ {\rm Hz} \ .
\end{eqnarray}
The first term in the equation above accounts for the remaining effect
of the atmosphere after eighty percent calibration is applied to the
data with the use of a water vapor radiometer \cite{Armstrong2006},
while the second term accounts for the effect of aperture averaging,
which causes a reduction in delay fluctuations on time scales less
than the antenna wind speed crossing time ($1$ to $10$ seconds)
\cite{Linfield1998}.

The {\it Transponder} entering into the $y$ measurement is responsible
for keeping the phase coherence between the incoming and outgoing
radio signals on the spacecraft.  Its performance depends on the
accuracy of tracking of the up-link signal by the phase locked loop,
and the noise floor and non-linearities of its electronic components
\cite{Riley_etal1990}. Frequency stability measurements of the Ka-Band
($32$ GHz) transponder flown onboard the CASSINI mission have 
resulted in the following one-sided power spectral density of the
relative frequency fluctuations
\begin{equation}
S_{TR} (f) = 1.6 \times 10^{-26} \ f \ \ {\rm Hz}^{-1} \ .
\end{equation}

The onboard {\it clock} provides the frequency and timing reference
for the onboard radio instrumentation, and identifies the stability of
the microwave signal transmitted to the ground. The space-qualified
clock presently under realization at the Jet Propulsion Laboratory
relies on a combined ``interplay'' between a local quartz oscillator
and a trapped Hg ions clock. The frequency of the hyperfine transition
made by the Hg ions is used for monitoring and correcting the
frequency of the quartz oscillator. This steering process takes place
on a typical time scale of about $10$ seconds, making then possible over
longer time scale to significantly improve the stability of the
resulting combined instrument. A frequency stability comparable to
that of the Deep Space Network ground clocks has already been
demonstrated with a prototype, showing an Allan standard deviation of
a few parts in $10^{-15}$ at an integration time of a few thousand
seconds\cite{Prestage_Weaver2007, Dick1990}.  The corresponding
one-sided power spectral density of the relative frequency
fluctuations of such a clock is given by the following expression

\begin{eqnarray}
S_{C_S} & = & 5.0 \times 10^{-27} \ \ \ \ {\rm Hz}^{-1} \ \ \ \ \
\ \ \ \ \ \ 
10^{-5} \le f \le 2.0 \times 10^{-2} \ \ \ \ \ {\rm Hz} \nonumber \\ 
& = & 2.5 \times 10^{-25} f \ \ {\rm Hz}^{-1} \ \ \ \ \ \ 
2.0 \times 10^{-2} \le f \le 2.0 \times 10^{-1} \ \ \ {\rm Hz}
\nonumber \\ 
& = & 10^{-26} f^{-1}  \ \ \ \ \ \ \ {\rm Hz}^{-1} \ \ \ \ \ \ 
2.0 \times 10^{-1} \le f \le 1 \ \ \ \ \ \ \ \ \ \ \ \ \ \ \ {\rm Hz}
\end{eqnarray}

The onboard {\it digital receiver} used in this method measures the
amplitude and phase of the uplink signal and telemeters that
information to the ground.

The frequency fluctuations of the receiver chain on board the
spacecraft is estimated to be entirely due to thermal noise because of
a simpler cabling system, and its performance being essentially
identical to that of the ground receiver. By assuming again a $34$
meter diameter beam waveguide antenna transmitting with an $800$ W
Ka-Band ($32$ GHz) amplifier to a spacecraft out to a distance of
$5.5$ AU, equipped with a $4$ meter diameter (HGA), and a system noise
temperature of $400$ Kelvin, we find the following one-sided power
spectral density of the frequency fluctuations of the onboard
electronics noise at Ka-Band \cite{Armstrong2006}
\begin{equation}
S_{EL_S} (f) = 7.2 \times 10^{-28} \ f^2 \ {\rm Hz}^{-1} \ .
\end{equation}

\end{document}